\documentstyle{l-aa}    % formato AA
\def\ltsim{\raise 2pt \hbox {$<$} \kern-1.1em \lower 4pt \hbox {$\sim$}}
\def\ltapprox{\raise 2pt \hbox {$<$} \kern-1.1em \lower 5pt \hbox {$\approx$}}
\def\gtsim{\raise 2pt \hbox {$>$} \kern-1.1em \lower 4pt \hbox {$\sim$}}
\def\gtapprox{\raise 2pt \hbox {$>$} \kern-1.1em \lower 5pt \hbox {$\approx$}}
\def\degrees{$^{\circ}$}

\def\skuno{\vskip 20pt}

\def\p0{\phantom{0}}

\def\ph1{\phantom{1}}
\begin{document}
\thesaurus{11.03.4 A~119; 13.18.1; 02.13.1; 11.09.3}
\title {The radio galaxies and the magnetic field in Abell 119}
\skuno
\skuno
\author{L. Feretti\inst{1} \and D. Dallacasa\inst{1} 
 \and F. Govoni\inst{2} \and G. Giovannini\inst{1,3} 
\and G. B. Taylor\inst{4} \and U. Klein\inst{5} }
\offprints{lferetti@ira.bo.cnr.it}
\institute{
Istituto di Radioastronomia del CNR, Via P. Gobetti 101, I-40129 Bologna, 
    Italy
\and
Osservatorio Astronomico di Padova, Vicolo dell'Osservatorio 5, I-35122
Padova, Italy
\and
Dipartimento di Fisica, Universit\'a di Bologna, 
I-40100 Bologna, Italy
\and
National Radio Astronomy Observatory, PO Box O,
Socorro, NM 87801 0387
\and Radioastronomisches Institut der Universit\"at Bonn, Auf dem H\"ugel 71,
D-53121 Bonn, Germany}

\maketitle

\begin{abstract}

We present new, multiwavelength Very Large Array observations of the
3 radio galaxies in the cluster Abell 119: 0053-015, 0053-016 and 3C~29.
The first two radio galaxies, which lie close to the cluster center,
show a narrow-angle-tailed structure, with many twists in the tails. 
The third radio source is located at 
the cluster periphery, and shows an undistorted FR\,I morphology.
All three radio sources are strongly polarized at the highest frequencies, and
all three show both more depolarization and higher Faraday Rotation
Measures with increasing proximity to the cluster center.
We interpret this polarization behaviour as induced by the magneto-ionized 
intracluster medium, whose magnetic field is estimated to be in the
range 5-10 $\mu$G.


\end{abstract}

\section{Introduction}

The intra-cluster  medium (ICM) in clusters of galaxies is known to
possess magnetic fields whose origin and properties are not yet well
known. The presence of cluster magnetic fields is demonstrated  by a)
the existence  of  diffuse  cluster-wide radio emission (radio halo)
as revealed in some clusters (e.g. Coma, see Giovannini et al. 1993,
and references therein), 
b) the detection of Inverse Compton hard X-Ray emission
(Bagchi et al. 1998, Fusco-Femiano et al. 1998),
c) the study of variations
of the Faraday Rotation of background sources shining through
different lines of sight across the clusters, d) the analysis of
Faraday Rotation gradients in extended sources embedded within the
cluster.

%The strength of the ICM magnetic fields covers a wide range of
%values. 
Kim et al. (1991) analyzed the Rotation Measure (RM) of radio
sources in a sample of Abell clusters and found that $\mu$G level
fields are widespread in the ICM, regardless whether they do or do
not have a strong radio halo.
Stronger magnetic fields, from about 5 up to the
extreme value of 30 $\mu$G (as in 3C~295, Perley \& Taylor 1991;
and Hydra A, Taylor \& Perley 1993) have been found in ``cooling flow'' 
clusters where extremely
large Faraday rotations have been revealed, suggesting that the
generation of very strong ICM magnetic fields may be connected with
the cooling flow process (Soker \& Sarazin 1990, Godon et al. 1998).
In  the Coma cluster, a magnetic field of about 6$h_{50}^{1/2}$ $\mu$G
was found by Feretti et al. (1995) from the
analysis of the rotation measure in the cluster radio galaxy NGC~4869.
This large value is  comparable to the magnetic field strength observed in
cooling flow clusters, and it is considerably larger than the
``equipartition'' field derived by the radio data of the diffuse radio halo
Coma C, permeating the Coma cluster center. 
%An implication of this result
%is that the  energy associated with the magnetic field is, 
%at least locally, comparable to the turbulent and thermal energy.

The ICM magnetic field can be tangled on scales much smaller
than the typical galaxy size.
Crusius-W\"atzel et al. (1990), studying the depolarization in 5 strong double
sources, found tangling on smaller scales (1-4 kpc). This is confirmed
by the results of Feretti et al. (1995) on the Coma cluster.

The knowledge of the properties of the large-scale magnetic fields in clusters
is important to study the cluster formation and evolution, and 
has significant implications for primordial
star formation (Pudritz \& Silk 1989). 
It has been suggested that strong fields can originate either by large
scale dynamo amplification (Ruzmaikin 1989) or by turbulence following a
cluster merger (Tribble 1993). 
These magnetic fields can be traced by studying the rotation measures
of radio sources located within or behind the cluster. 

\begin{table*}
\caption{VLA Observing Log}
\begin{flushleft}
\begin{tabular}{llllll}
\hline 
\noalign{\smallskip}
Name &  Frequency  & Bandw & Config. & Date & Duration  \\
     & MHz  &   MHz        &         &    &   Hours  \\
\noalign{\smallskip}
\hline
\noalign{\smallskip}
 A~119    & 1365/1515$^*$ & 12.5 & B & Jan96       & 8   \\
          & 1365/1515$^*$ & 12.5 & C & Feb96       & 3   \\
 0053-015 & 4835/4885     & 50   & B & Dec95-Jan96 & 6   \\
	  & 4835/4885     & 50   & C & Feb96-Mar96 & 2.1 \\
          & 8415/8465     & 50   & B & Dec95-Jan96 & 6   \\
          & 8415/8465     & 50   & C & Feb96-Mar96 & 2.1 \\
	  & 7815/8165     & 50   & C & Nov94       & 0.7 \\
	  & 8515/8885     & 50   & C & Nov94       & 0.6 \\
 0053-016 & 4835/4885     & 50   & B & Dec95-Jan96 & 6   \\
          & 4825/4885     & 50   & C & Feb96-Mar96 & 2.1 \\
	  & 7815/8165     & 50   & C & Nov94       & 0.7 \\
	  & 8515/8885     & 50   & C & Nov94       & 0.6 \\
          & 8415/8465     & 50   & B & Dec95-Jan96 & 4   \\
          & 8415/8465     & 50   & C & Feb96-Mar96 & 2.1 \\
 3C~29 & 4835/4885     & 50   & C & Mar96       & 0.7 \\
          & 8415/8465     & 50   & C & Mar96       & 0.7 \\
\noalign{\smallskip}
\hline
%\noalign{\smallskip}
\label{olog}
\end{tabular}
\end{flushleft}
$^*$ Higher frequency IF not used because of interferences
\end{table*}

We examine here the cluster Abell 119, which is characterized by the
presence of three extended radio galaxies, located at different
projected distances from the cluster center. Two sources, 0053-015 and
0053-016, show a narrow-angle-tailed (NAT) structure of $\sim$5\arcmin~ in
size, and are projected close to the cluster center. The
third source, 3C~29 (0055-016), is a  typical FR~I,
extended about 2.5\arcmin~ and located  at the cluster periphery.
All these sources are highly polarized, therefore they are suitable for
a study of the rotation measure. Moreover, their presence at different
cluster locations is crucial to  
derive information about the magnetic field
strength  and structure in the whole cluster.
The cluster A~119 (z=0.0441) is of richness class 1, and is
classified as BM type II-III. The first ranked  galaxy
is UGC~579,  classified as a cD galaxy 
(Postman \& Lauer 1995, Saglia et al. 1997). 
%A weak cooling flow of 23 M$_{\odot}$ yr$^{-1}$ is reported by  
%Edge et al. (1992). 
According to Peres et al. (1998), no cooling flow is present in this 
cluster. 

We use a Hubble constant H$_0$ = 50 km s$^{-1}$ Mpc$^{-1}$ throughout
this paper. At the distance of A~119, 1 arcsec corresponds to 1.18
kpc. 

\section {Observations and data reduction}

The data presented here were obtained with the Very Large Array (VLA)
at multiple frequencies, 
in the B and C configurations, as given in Table \ref{olog}. 
The source  3C~48 was used as a primary flux
density calibrator. The phase calibrator was the nearby point source
0056-001, observed at intervals of about 20 minutes, 
while the polarization position angle calibrator was
 3C~138. The instrumental polarization of the antennas
was corrected using the secondary calibrator 0056-001, which was
observed over a wide range of parallactic angles.
A single pointing for the whole cluster at 1.365/1.515 GHz allowed us to
obtain images of  the two tailed radio galaxies
%individual radio galaxies, 
with negligible bandwith smearing.  The radio galaxy 3C~29, offset
by $\sim$20\arcmin~ from the pointing position, suffers of a reduction in
peak response of about 50\%.
The data at 1.515 GHz  were seriously affected by a strong time-variable
interference, which limited the sensitivity achieved in the 
image; therefore they were not used in the analysis.
At the higher frequencies each source was observed individually.
The data were reduced with the Astronomical Image Processing System
(AIPS), following the standard procedure: Fourier-Transform, Clean and
Restore. Self-calibration was applied to minimise the effects of
amplitude and phase variations.
All the images were corrected for primary beam attenuation.

Images of the Stokes parameters I,
Q and U were produced at each frequency, with different resolutions,
using the AIPS task IMAGR. The restoring beam was typically a circularly
symmetrical Gaussian.
The images of the polarized intensity $ P =  (Q^2 + U^2)^{1/2}$, 
the degree of polarization m = P/I and the position angle of polarization
$\theta = 0.5 \tan^{-1} (U/Q)$ were derived from the I, Q and U images.
The P maps were corrected for the positive Ricean bias
due to the combination of two noisy quantities in quadrature
(Wardle \& Kronberg 1974).  
 The errors deriving from off-axis primary beam polarization 
are always less than 1\% in our sources, except in 3C~29
at 1.4 GHz. At its distance from the 
phase center ($\sim$ 20\arcmin) the error on the polarization percentage is 
estimated to be within $\sim$3.5\% (Napier 1989).

\vskip 5pt
\section { Total intensity and polarization images }

\subsection {0053-015}

The radio galaxy 0053-015 is classified as a NAT 
source, with a total extent of $\sim$5\arcmin. 
It was mapped at  1.4 GHz by  O'Dea \& Owen (1985), who found a quite
irregular structure with asymmetric jets. The large scale image and
the polarization information, obtained with the Effelsberg radio Telescope, 
were presented by Mack et al. (1993).  The source is highly polarized 
at  10.7~GHz, the fractional polarization reaching about 30\%, and a
magnetic field orientation parallel to the tail direction.

The overall source structure is easily visible at 1.4 GHz, at a
resolution of 3.5\arcsec~ (Fig. \ref{Cont15l}). The total extent of the 
source mapped at this frequency with the VLA is  about 
5\arcmin, which is comparable to the extent found from the
single dish image (Mack et al. 1993). The unresolved core, located in
%and accounts for about 10\% of the total flux density at 4.9 GHz.
the upper western region (C in Fig. \ref{Cont15l}), 
is very  bright in the images at all frequencies. 
The two short symmetric jets emanating from
the core are oriented in the NE-SW direction. 
The NE jet bends by about 90\degrees~ to the south
to form a low surface brightness lobe, which expands and merges with the
emission originating from the other jet. 
The long tail, notably rich in wiggles, smoothly fades into the noise.

The highest resolution image at 4.9 GHz, presented in
Fig. \ref{Cont15c}, shows in great detail the structure of the jets
and lobes. The SW jet is well collimated for about
20\arcsec, then it widens suddenly forming a well defined 
lobe, characterized by several bends and wiggles. The brightness of the
western lobe shows a  maximum at $\sim$50\arcsec~ from the nucleus,
and decreases progressively further out.
The NE jet is straight at the beginning, then it sharply
bends by 90\degrees, and forms a bright large lobe.
This jet is slightly brighter than the opposite one, and is easily
distinguishable for about 20\arcsec. Unlike the SW jet,
it is surrounded by a cocoon of lower brightness.
The lobe shows a squared corner toward the east, then  bends and
twists, before quickly fading  into the low brightness tail. 
Although the structure is very skewed, the properties of the two jets
and lobes show some symmetries:  similar length of the jets, similar
distance from the nucleus of the two regions of maximum brightness in
the lobes, and finally the winding structure of the tails.
The jet structure is enhanced in the 8.4 GHz image
(Fig. \ref{Cont15x}), where the low brightness lobes are mostly
resolved out.

The source is strongly polarized at 8.4 GHz and 4.9 GHz, with similar
values of the polarization percentage. In the image
with 3.5\arcsec~ resolution, the fractional polarization is $\sim$4\%
in the nucleus, then it oscillates between 10\% and 35\% along the two
lobes, with values up to 50\% at  the lobe boundaries.
At higher resolution (HPBW=1.5\arcsec), the
degree of polarization is only slightly higher.
At 1.4 GHz,  the polarization percentage drops to values always below
10\%, with irregular variations along the structure. The average 
depolarization, defined as the ratio between the polarization percentage
at 1.4 GHz, m$_{1.4}$,  and that at 4.9 GHz, m$_{4.9}$, is of $\sim$0.2.
%The magnetic field orientation is circumferential to the edges of the
%emission region.

\subsection {0053-016}

The NAT radio galaxy 0053-016 was imaged  at  1.4 GHz by  O'Dea \&
Owen (1985), who found a total size similar to that of 0053-015 
($\sim$5\arcmin), and a quite symmetric and regular structure.
Despite the similarity of the structure of the two tails, 
the fractional polarization
detected at 10.7~GHz by Mack et al. (1993) is very asymmetric. Significant
polarized flux up to a level of 20\%  is found by these 
authors in the 
western source region, and in the outermost tail, while the eastern
region is unpolarized. This structure in polarization has been interpreted
by Mack et al. (1993) as  due to  depolarization within the observing
beam of $\sim$ 1\arcmin.
%The magnetic field orientation is again parallel to the source major axis.

In the 1.4 GHz image obtained with 3.5\arcsec~ resolution
(Fig. \ref{Cont16l}), the head of the source is resolved in two
symmetric jets, bent backward, and without clear evidence
of the nucleus (labelled as C in the figure). 
The jets progressively widen, and  follow a helical
path, before merging into a low surface brightness tail.  
In the high resolution images at higher frequencies, 
given in Figs. \ref{Cont16c} and \ref{Cont16x}, the core is 
easily visible between the two symmetric opposite jets, initially
oriented in the SE-NW direction. Both jets are curving rather
continuously,  but the western jet shows a slightly sharper bend at
the northest  point of the source, similar to  0053-015.
The two jets show an increasing width, with regions of high
brightness, and an extraordinarily twisted structure, up to a
projected distance of about 100\arcsec~  from the core. Beyond this
distance, there is a remarkably straight and very collimated region in
the western jet, about 20\arcsec~ long (around position RA = 00$^h$
53$^m$ 25.5$^s$, DEC = --01\degrees~ 37\arcmin~ 25\arcsec), after
which  the jet seems to disrupt, forming a wider low surface
brightness tail which is sharply bent to the south. 

The core shows a polarization percentage of 6\%. 
Both jets are higly polarized at 4.9 GHz and 8.4 GHz. 
At 3.5\arcsec~ resolution,
the fractional polarization at both frequencies oscillates between 5\% and 25\%
for about 100\arcsec, with peaks in the bends.
 In the straight narrow region
of the western jet, the polarization percentage increases to $\sim$60\%.
It seems clear that the high fractional polarization in this region is
responsible for the polarization asymmetry detected in the Effelsberg data
at 10.7 GHz (Mack et al. 1993). At the higher resolution of 1.5\arcsec~
the polarization percentage is higher, thanks to the lower beam depolarization.
At 1.4 GHz, the source is  depolarized, with m$_{1.4}$/m$_{4.9} \sim $0.5.

%The magnetic field direction is parallel to the 
%jet trajectory, following its helical structure, and circumferential to the 
%edges of the structure.

\subsection {3C~29 (0055-016)}

This radio galaxy has been previously observed with the VLA by
Morganti et al. (1993) and is classified as FR~I. It is considered 
to be a rather isolated object (Fasano et al. 1996), in agreement with
its peripheral cluster position.

The image of this galaxy at 4.9 GHz  with 5\arcsec~ resolution is 
presented in Fig. \ref{Cont3c}. It shows the central nucleus and two opposite
symmetric straight jets, approximately oriented in the N-S
direction. The cocoon of radio emission around the jets is typical of
FR~I sources (De Ruiter et al. 1990). 
In the image at 8.4 GHz, with the highest
resolution  allowed by our data (Fig. \ref{Cont3cx}), the low
brightness structure is still visible, and the jets are 
enhanced. The structure of the source magnetic field is consistent 
with that of FR~I's. The fractional polarization at 4.9 GHz and 8.4 GHz
is  about 45\% in the northern jet,  25\% in the southern one, and 
reaches 50\% at the lobe boundaries. At 1.4 GHz the source is still well
polarized, with 
little depolarization (m$_{1.4}$/m$_{4.9} \sim$ 0.8). 

%The magnetic field direction is transversal in the jets and circumferential
%in the outer low brightness structure.

\section {Properties of the cluster radio galaxies}

\subsection { Spectra of the tailed radio galaxies}

The point-to-point spectral index along the source structure of the
two NAT's 
%up to a distance of $\sim$2\arcmin~ from the core 
was derived using the images at 1.4 GHz, 4.9 GHz and 8.4 GHz,
obtained with the angular resolution of 3.5\arcsec~ from data with
matched uv-coverage. In both sources the spectrum of the core is flat
with $\alpha \sim$ 0. At the jet bases the spectral indices are about
0.5-0.6, and evidence of spectral curvature is found with increasing
distance from the core, along the tails. The spectral indices at a
distance of $\sim$2\arcmin~ are similar in the two sources, with
$\alpha^{1.4}_{4.9}$ $\sim$ 0.8-0.9 and  $\alpha^{4.9}_{8.4}$ \gtsim 2. 
%The curvature in the spectrum could be related to the magnetic
%field strenght changing across the source, but the dominant effect is
%likely to be the ageing of the radiating electrons.
%We then assume that this curvature is the result of synchrotron losses from an

 A curvature in the emission spectrum is expected as a result
of synchrotron energy losses from an 
ensemble of electrons with an initial power-law energy distribution, 
and appears at a critical frequency $\nu_{\rm c}$, related to the 
electron age. The shape of the expected synchrotron spectrum can be computed
analytically as a function of the critical frequency, the electron energy
distribution index  and the evolution of the electron pitch-angle
distribution with time (Pacholczyk 1970). The model of Kardashev-Pacholczyk
(KP) is obtained in the case that electrons maintain the same pitch angle
throughout their radiative lifetime (Kardashev 1962). The model of
Jaffe-Perola (JP) assumes that there is a redistribution of electron pitch
angles on short time scales in comparison with their radiative lifetimes, due
to their scattering on magnetic field irregularities (Jaffe \& Perola 1973). 

\begin{table*}[t]
\caption{Optical data on radio galaxies}
\begin{flushleft}
\begin{tabular}{llllllll}
\hline 
\noalign{\smallskip}
Name & Other & RA \ \ (B1950)  & DEC & m$_R$ &  v$_{hel}$ 
 & $\mid \Delta$ v $\mid$ & Proj. Dist. \\
   & & h \ m \ s  & \degrees~ \ \arcmin~ \ \arcsec~ & & km s$^{-1}$ &
 km s$^{-1}$ & \ \ \arcmin~ \ \ \ kpc \\
\noalign{\smallskip}
\hline
\noalign{\smallskip}
 0053-015 & UGC~583 & 00 53 52.3  & -01 31 59 & 12.93 & 11456 &  1697 &  
2.4\ \ \ \ 170 \\
%          & R: & 00 53 52.22 & -01 31 58.2 &       \\
 0053-016  & -- & 00 53 29.3  & -01 36 17 & 13.89 & 12795 &  415 &  
6.4\ \ \ \ 453 \\
%          & R: & 00 53 29.34 & -01 36 17.4 &       \\
 3C~29  & UGC~595 & 00 55 01.6  & -01 39 40 & 12.80 & 13491 &  252 &  
21.4\ \ \ 1515 \\
%          & R: & 00 55 01.57 & -01 39 39.5 &       \\
   \noalign{\smallskip}
\hline
%\noalign{\smallskip}
\label{galad}
\end{tabular}
\end{flushleft}
Caption. Columns 1 and 2: source name; 
Columns 3 and 4: coordinates of the galaxies from Colless at al. (1993);
Column 5: red apparent magnitudes (from Saglia et al. 1997); Column 6:
heliocentric velocities from Way et al. (1998); Column 7: proper
velocities with respect to the average velocity of the cluster;
 Columns 8 and 9: projected angular and
linear distance from the cluster center (given in Table \ref{clusd}). 
\end{table*}

\begin{table*}[t]
\caption{Cluster X-ray parameters}
\begin{flushleft}
\begin{tabular}{lllllll}
\hline 
\noalign{\smallskip}
Name &  RA \ \ (B1950) & DEC  & r$_c$ & $\beta$ & n$_0$ & T \\
 & h \  m \ s  & \degrees~ \ \arcmin~ \ \arcsec  & kpc &   & cm$^{-3}$ & keV \\
\noalign{\smallskip}
\hline
\noalign{\smallskip}
 A~119 &  00 53 43.5 & --01 30 59 
  & 378 & 0.56 & 1.18$\times$10$^{-3}$ & 5.6 \\
   \noalign{\smallskip}
\hline
%\noalign{\smallskip}
\label{clusd}
\end{tabular}
\end{flushleft}
Caption. Columns 2 and 3: coordinates of the X-ray centroid (from Peres et
al. 1998); Column 4: core radius; Column 5: ratio of the 
galaxy to gas temperature; Column 6: central gas density; 
Column 7: temperature.
\end{table*}

We obtained the spectra of our sources between 1.4 GHz and
8.4 GHz, at several locations along the tails up to $\sim$3\arcmin~ from
the radio nucleus in 0053-015 and up to a distance of  $\sim$2\arcmin~ 
in 0053-016. 
In the outermost points, only  upper limits to the 
8.4 GHz brightness were derived. These spectra were then 
%The spectra of our sources at various locations along the tails were 
fitted using both the KP and JP models, since we do
not have observational evidence in favour of any of them.
The fitting procedure was developed by Murgia \& Fanti (1996) 
 and applied using the same approach as in Feretti et al. (1998).
We used an initial electron energy distribution function of the kind 
N(E)dE $\propto$ E$^{-2}$dE, corresponding to a synchrotron
spectral index $\alpha$=0.5

Good fits to the observed spectra were  obtained by both the KP and JP 
model  within $\sim$ 1\arcmin~ from the core, while at further distances
the measured spectra are steeper than expected from the models. 
This could possibly derive from the limited 
%a progressivly decreasing 
sensitivity to low surface brightness structure in the 8.4 GHz image.
 We note also that non-uniformities in the particle and field
distributions in the sources can complicate the interpretation of
spectral curvature (Wiita \& Gopal-Krishna 1990, Eilek \& Arendt 1996).

From the values of the critical frequency and of the magnetic
field, it is possible to get information on the lifetime of radiating
electrons suffering synchrotron and inverse Compton losses. 
As the magnetic field strength, we used the equipartition value
estimated at the various locations along each source. The calculation
of the equipartition magnetic field involves a number of assumptions 
(Pacholczyk 1970). We assumed 
a cylindrical geometry, a volume filling factor of 1,
upper and lower frequency cutoffs of 10 GHz and 10 MHz, and equal energies
in the relativistic electrons and protons.
In the source 0053-015, the value of the equipartition magnetic field 
decreases with distance from $\sim$7 $\mu$G in the inner
points to $\sim$ 2.5 $\mu$G at $\sim$3\arcmin, while 
the  critical frequency $\nu_{\rm c}$ decreases from \gtsim 30 GHz to
$\sim$ 7 GHz. This leads to electron ages in the range 1-5 $\times$
10$^7$ yr, and to an 
average projected bulk velocity of $\sim$ 4500 km s$^{-1}$ in both tails.
In 0053-016, the equipartition magnetic field close to the
nucleus is $\sim$8.5 $\mu$G, and at $\sim$2\arcmin~ from the
core is 3.8 $\mu$G in both tails. The critical frequency
ranges from $\sim$ 30 GHz to $\sim$ 10 GHz, and the derived ages are about
1-4$\times$10$^7$ yr. This implies an average electron projected bulk
velocity of $\sim$ 4000 km  s$^{-1}$.

 The relativistic electron velocities are higher than expected, as 
we would expect that the particles diffuse at their typical sound speed
or that  their motion reflects the proper motion of the parent galaxy
within the cluster (see next Subsection). 
We have also considered the possibility that the magnetic field which
enters in the computation of the age is not the equipartition value.
We have noted, however, that the equipartition magnetic fields in the source
extremes are close to the values which give the maximum lifetimes
(Van der Laan \& Perola 1969), therefore the use of significantly
different values of the magnetic field would make the velocity still higher.

\subsection {Optical data}

The host galaxies of the three radio sources studied here 
are all classified as ellipticals, with no particular
features. The parent galaxy of 3C~29 shows round optical contours
(Smith \& Heckman 1989) also confirmed by the HST image 
(Zirbel \& Baum 1998). 

The dynamical analysis of the structure of A~119 has been recently
presented by Way et al. (1998), who give  velocities for 153 cluster 
galaxies and derive an average cluster velocity of 13228
km s$^{-1}$, and a velocity dispersion of 778 km s$^{-1}$. 
They also suggest the possible presence of subgroups.
%cluster actually
%consists of a main group of 125 members
%and two smaller groups of 11 and 17 members, respectively. 
%The radio galaxies 0053-016 and 3C~29 are members of the main
%cluster, whose heliocentric velocity is v$_{hel}$=13248 km s$^{-1}$,
%with $\sigma_{vel}$ = 472 km s$^{-1}$,  while the
%radio galaxy 0053-015 belongs to a small subgroup of lower average
%velocity (v$_{hel}$=11699 km s$^{-1}$, $\sigma_{vel}$ = 291 km s$^{-1}$).  
In Table \ref{galad}, the positions of the optical parent galaxies
are given, together with their radial heliocentric velocities.
The galaxy proper velocities  have
been calculated with respect to the average cluster velocity.
% of the 
%corresponding subgroup. 
From the proper velocities of the three radio galaxies it
is evident that the motion along the line of sight is
quite low in 0053-016 and 3C~29. We can argue that the peripheral source
3C~29 has a genuinely low proper motion, in agreement with its
undistorted structure, while the NAT 0053-016 has a 
predominant velocity component in the plane of the sky.
In the previous subsection we found that the bulk velocity of radiating
particles in the NAT galaxies is about 4000 - 4500 km s$^{-1}$.
It is likely that part of this velocity reflects the proper motion of
the parent galaxy within the cluster; however, it seems too high to 
fully account for it. The high velocities found for the electrons
in the tails can be interpreted as due to: 
a) projection effects; b) the existence of reacceleration processes, which 
prevent the electron spectral index steepening 
and allow the electrons to extend their lifetimes; c) the presence 
of genuine bulk streaming motion in the radio
emitting plasma;  d) the existence of
motion of the cluster medium (Loken et al. 1995);
e) a non-uniform magnetic field,  which affects the estimate of the
critical frequency. 
Similar results have been found in a small sample of
tailed radio galaxies by Feretti et al. (1998).

\section {Comparison of X-ray and radio images}

The cluster A~119 has been the target of X-ray observations with the
ROSAT PSPC, for a total exposure time of 15203 sec. 
The data were analyzed by Cirimele et al. (1997), 
who fitted the X-ray brightness with a  hydrostatic 
isothermal model of the form
$$S(r) =S_0 (1 + r^2/r^2_{\rm c})^{-3 \beta + 0.5} \eqno(1)$$
 where S$_0$ is the central surface brightness, $r_{\rm c}$ is the
core radius, and  $\beta$ is the ratio of the galaxy
to gas temperature. They derived the parameters given in 
Table \ref{clusd}. 
The gas temperature in the Table is from  Markevitch et al. (1998),
who obtained a temperature map from ASCA data, and concluded that the
temperature profile of this cluster is nearly constant.
We retrieved the data from the public ROSAT archive
and produced an image by binning the photon
event table in pixels of 15\arcsec, and by smoothing the map
with a Gaussian of $\sigma$ = 45\arcsec. 
In Fig. \ref{Radx} the X-ray brightness distribution is overlaid onto the
radio emission from the three cluster radio galaxies. 
The X-ray brightness distribution is rather irregular and asymmetric, with an 
extension in the northern region. The 
centroid of the X-ray emission (given in Table \ref{clusd})
 is consistent with 
%is displaced by 36\arcsec~ from  
the position of the cD galaxy (Peres et al. 1998).
%was obtained by a gaussian fit to the
%innermost region, and is 00 53 42, -01 31 21
The three radio galaxies are  all embedded within the X-ray gas,
at the projected distances given in Table \ref{galad}. The two NAT's
are located within 1.2 core radii from the cluster center,
while 3C~29 is at 4 core radii, 
%Finally, X-ray emission is detected out to the distance of 3C~29, which 
%is found 
and coincides with a weak X-ray local enhancement.
It is difficult to do an effective comparison between the radio 
emission and the X-rays, given the lower resolution  of the ROSAT PSPC
data. 
%However, there is  evidence that the tail of
%0053-016 is developing in a region of lower hot gas density; furthermore,
%the southern bend of the western tail of 0053-016, at about
%120\arcsec~ from the core, seems to avoid a region of enhanced X-ray
%emission. On the other hand, 0053-015 appears to be constrained
%between the cluster center and a secondary peak of X-ray emission to
%the south, with no distinct local variation of the hot gas density. 
It is worth noting that the tails in both 0053-015 and 0053-016
have very similar orientations, even though the density gradient, related
to the X-ray contours, is quite different.  This could be related
with the presence of merger-induced bulk motion of the intergalactic
medium, as suggested by Bliton et al. (1998).

\section { Rotation measure structure}

We obtained images of the rotation measure  in the radio galaxies
under study by combining the suitable polarization data available to us.
For the two tailed radio galaxies, 0053-015 and 0053-016, 
we used the maps at the 5 frequencies
4.835, 7.815, 8.165, 8.515 and 8.885 GHz, with 3.75\arcsec~ resolution. 
In principle, the addition of the 1365 GHz could
provide a larger wavelength sampling, but the polarization of the two 
sources is too low at this frequency to be useful.

Following the definition  $\theta_{\lambda} = \theta_i + 
RM \lambda^2 $, where $\theta_i$ is the intrinsic position angle of the 
polarization vector and $\theta_{\lambda}$ is the position angle at
wavelength $\lambda$, the value of the RM 
was computed by linear fitting of
the polarization  angle as a function of $\lambda^2$.
The pixels in which the uncertainty in the polarization angle exceeds
20\degrees~ were blanked.  We note that only a few pixels near the
source edges have such large uncertainties. 
The fitting algorithm provides a weighted least-squares fit, allowing
for an ambiguity of $\pm$n$\pi$ in each polarization position angle.

The data of both sources are well fitted by a 
$\lambda^2$ relation. 
 The typical error in the resulting values of the
rotation measure is $\sim$30 rad m$^{-2}$.
The image of RM in 0053-015 is  presented in Fig. \ref{rm15}.
The values of RM range between --350 rad m$^{-2}$ and +450 rad m$^{-2}$,
and show fluctuations on small scales. 
The distribution of the values
is evident from  the histogram in Fig. \ref{rmhi15}.
It is non-Gaussian, with the peak close to zero.
The average RM of the whole source is $<$RM$>$ = + 28 rad m$^{-2}$,
with a dispersion of $\sigma_{RM}$ = 152 rad m$^{-2}$.

The image of RM in 0053-016 is  presented in Fig. \ref{rm16}.
The values of RM range between --300 rad m$^{-2}$ and +200 rad m$^{-2}$,
as displayed by the histogram of Fig. \ref{rmhi16}.
The distribution of RM shows two peaks, both negative.
The average RM of the whole source is $<$RM$>$ = --79 rad m$^{-2}$,
with a dispersion $\sigma_{RM}$ = 91 rad m$^{-2}$.

For 3C~29 we obtained the rotation measure by comparing the images at
1.365 GHz, 4.885 GHz and 8.440 GHz, with 5\arcsec~ resolution, given 
that the source is highly polarized also at the low frequency. 
The RM in 3C~29 is rather uniform over the whole 
source, with values between --30 and +30 rad m$^{-2}$
(see Figs. \ref{rm3c} and  \ref{rmhi3c}). 
The average value is $<$RM$>$ = +4 $\pm$ 13 rad m$^{-2}$.
The RM is therefore consistent with zero. 

All the values of RM obtained in this section are likely to be genuinely
related to the sources under study. The Galactic contribution to RM in
the region of A~119 is in fact expected to be only $\sim$1 rad m$^{-2}$
(Simard-Normandin et al. 1981), and therefore negligible.

\section{Discussion}

\subsection{Evidence for cluster magnetic fields }

Faraday rotation in a radio source can be produced if the line
of sight crosses a region of mixed magnetic field and ionized thermal gas
(external Faraday rotation).
In this case, the polarization angle obeys a $\lambda^2$
law over an angle larger than $\pi$/4, and 
the relation between the RM, the gas density n$_e$,
and the magnetic field along the line of sight $B_{\parallel}$ is
given by 
$$ RM=812 \int B_{\parallel} n_e dl \ \ \ \  {\rm rad~m^{-2}} \eqno(2)$$
where $B_{\parallel}$ is measured in $\mu$G, $n_e$ in cm$^{-3}$ and
$dl$ in kpc. 
This is the case for radio sources in clusters, if the hot intracluster gas
is magnetized.  The RM distribution of cluster radio galaxies can therefore
be used to derive information on the magnetic field along the line
of sight.

\begin{table}[t]
\caption{Summary of RM and depolarization data}
\begin{flushleft}
\begin{tabular}{llllll}
\hline 
\noalign{\smallskip}
Name & $\mid$RM$_{max}\mid$ & $<$RM$>$ & $\sigma_{RM}$ & DP & Dist \\
     & rad m$^{-2}$ &  rad m$^{-2}$  & rad m$^{-2}$ &  & r/r$_c$ \\ 
\noalign{\smallskip}
\hline
\noalign{\smallskip}
 0053-015 & 450 & +28 & 152 & 0.2 & 0.45 \\
 0053-016 & 300 &  --79 & 91 & 0.5 & 1.20 \\
 3c~29 &  30 & +4 & 13 & 0.8 & 4.01 \\
\noalign{\smallskip}
\hline
%\noalign{\smallskip}
\label{polar}
\end{tabular}
\end{flushleft}
Caption. Column 1: source name; Column 2: maximum absolute value of RM; 
Column 3: average value of RM; Column 4: RM dispersion; 
Column 5: average depolarization defined
as m$_{1.4~GHz}$/m$_{4.9~GHz}$; Column 6: distance from the cluster
center in units of core radii.
\end{table}

In the case of a tangled magnetic field, 
with cells of uniform size, same  strength, and random 
orientation, the observed RM along any given
line of sight will be generated by a random walk process, and the
distribution of RM results in a Gaussian with zero mean, and the
dispersion related to the number of cells along the line of sight. The
source will also depolarize at long wavelength, if the external
Faraday screen is not fully resolved by the observing beam.

The good $\lambda^2$ fits to the polarization angle favour the
interpretation that external Faraday rotation is the dominant mechanism
in the present sources. 
%This is consistent with recent results in the
%literature (ref.). 
In Table \ref{polar} we summarize the results for the present radio
galaxies. The most striking result is the trend of RM dispersion and
depolarization with distance from the cluster center. The innermost
source, 0053-015, has the largest $\sigma_{RM}$,  the highest
absolute values of RM, and the highest depolarization at long
wavelengths. The source 0053-016, located just beyond 1 core radius
from the cluster center, still shows high values of RM, but lower than
in 0053-015, and also the depolarization is lower. Finally, the
peripheral source 3C~29 shows little RM and little depolarization.
This result points to the interpretation that the external Faraday
screen is the same for all 3 sources, i.e. it is the intergalactic medium
in A~119, which plays different roles according to how much
magneto-ionized medium is crossed by the polarized emission. 
%Although we can determine only the $projected$ distances of the radio
%galaxies from the cluster center, the indication from the RM data is that
 This is consistent with the two NAT's being really located in 
the inner cluster
region, and not simply projected onto it.
As suggested by Tribble (1991), unresolved external RM fluctuations
produce a fall-off of the polarization degree with $\lambda$.
A consistent picture is thus that the structure of the intergalactic 
magnetic field is tangled on small scales, this accounting for the
observed depolarization. 
From the polarization degree of 0053-015 and 0053-016 (see Sect. 3),
there is evidence that the 3.5\arcsec~ observing beam does not fully
resolve the screen. Thus, we can argue that the scale of tangling of the
magnetic field is $<$4 kpc. Moreover, field reversals must take place.

The indirect detection of 
the magnetic field associated with the intergalactic medium of
A~119 is an important result, since so far a significant
intergalactic magnetic field has been only found at the center of 
clusters with strong cooling flows (Ge \& Owen 1993, Taylor et al. 1994).
Moreover, direct evidence of a cluster magnetic
field is provided in the few clusters containing a radio 
halo (see e.g. Feretti \& Giovannini 1996). 
The magnetic field present in A~119 is spread on a size larger at least
than one cluster core radius.
The existence of a magnetic field component in the
intergalactic medium therefore seems to be a common feature in
clusters of galaxies. 

\subsection{Strength of cluster magnetic fields }

The determination of the strength of the magnetic field in A~119
depends on the model 
for  the X-ray gas distribution and on several assumptions, including
the RM structure function, and the field structure.

In the simplest case that a constant magnetic field fills the whole
cluster, using the central gas density and 
a core radius as the path-length, Eq. (2)
requires $B_{\parallel}$ = 1.2 $\mu$G.
We have found, however, that the magnetic field is likely to be
tangled on scales of the order of a
few kpc. In this case, the intensity of magnetic field is larger by the square 
root of the number of cells crossed by the line of sight. 

A more  realistic approach is to assume a magnetic field in randomly oriented 
 cells of uniform size and strength, and a gas 
density distribution given by the hydrostatic isothermal model.
The gas density $n_e$ has the functional form 
$$n_e(r) =n_0 (1 + r^2/r^2_{\rm c})^{-3 \beta/2} \eqno(3)$$
where the parameters have the same meaning as in Eq. (1).
In this case, the RM dispersion at 
different projected distance from the cluster center
was evaluated by Felten (1996) by solving the integral of Eq. (2):
$$ \sigma_{RM}= {{K B_{\parallel} n_0  r_c^{1/2} l^{1/2} }\over
{(1+r^2/r_c^2)^{(6\beta -1)/4}}} \sqrt {{\Gamma(3\beta-0.5)}\over{\Gamma
(3\beta)}}   \eqno(4) $$
where $\Gamma$ is the Gamma function, 
$r_c$ is the core radius in kpc, and $l$ is the size of
each cell in kpc;  the central gas density $n_0$ is in cm$^{-3}$ 
and $B_{\parallel}$ is in $\mu$G. The constant 
$K$ depends on the integration path over the gas density distribution:
$K$ = 624, if the source lies completely
beyond the cluster, and $K$ = 441
if it is as distant from the observer as the cluster center.
For the cluster A~119, using $K$ = 441
%Assuming the latter case to be suitable for the present galaxies,
and the gas parameters given in Table \ref{clusd}, 
%and a magnetic field cell of 2 kpc, 
the previous formula becomes
$$ \sigma_{RM} \approx {{10.2 B_{\parallel} l^{1/2}}\over
{(1+r^2/r_c^2)^{0.59}}} \eqno(5) $$
The values of $\sigma_{RM}$ obtained for the 2 NAT's of A~119
fit fairly well in the model described above and 
are consistent with a magnetic field strength in the range 9.6-13.6 $\mu$G,
for a cell size between 2 and 4 kpc.

The RM of the peripheral source 3C~29 is also  consistent with
the previous model.  However, the magnetic field probably cannot
be constant over 4 core radii ($\sim$ 1.5 Mpc), otherwise the magnetic
pressure would exceed the thermal pressure in the outer parts of the
cluster. So it is reasonable to believe that the very low RM 
of 3C~29  derives not only from the low density of the foreground
medium, but also from a weaker magnetic field in this region.

The model used to derive the magnetic field strength
is very simplified:
first, the two radio galaxies can be at different locations
along the line of sight; second, the cluster shows possible substructures
both in optical and  in X-rays, and therefore the model used for the
gas density distribution may be inaccurate; third, the magnetic
field structure could be more complicated than assumed.
We cannot exclude the possibility that the cell
size of the magnetic field in the region 
around the radio galaxies would be smaller than that of 
the the cluster-wide magnetic field,
due to the interaction between the radio sources themselves and the
intergalactic medium. 
With cells of 10-20 kpc, the magnetic field derived from Eq. (5) is
4.5-6.1 $\mu$G.

Even allowing for  the uncertainties related to the previous computations,
the observational evidence favours
the existence of a strong magnetic field  in the
intergalactic medium of A~119, over a scale larger 
than the cluster core radius. 
A plausible range for the magnetic field strength
is about 5-10 $\mu$G.

\section { Conclusions}

Three radio galaxies belonging to the cluster Abell 119, located at 
different projected distance from the cluster center, are studied
in total intensity and polarization. The two radio galaxies closer 
to the cluster center show similar NAT structure,
while the most peripheral radio galaxy is a classical FR\,I source.

The spectra of the NAT radio galaxies progressively steepen with the
distance, as expected under the effect of synchrotron energy losses.
The electron lifetimes inferred from the spectral curvature lead to 
drift velocities of $\sim$4000-4500 km s$^{-1}$. 
 Still higher values are obtained if the magnetic field is different 
from the equipartion value. Such large values of the velocities
can derive from projection effects, or from the existence of 
reacceleration processes within the tails, or bulk motions along the 
tails  and/or in the integalactic medium. We also note that 
non-uniformities in the magnetic field can affect  the interpretation 
of the spectral curvature. 

Large Faraday Rotations are found in the NAT's. 
The magnitude and
scale of the RM is consistent with a magnetic field embedded in the
hot X-ray emitting cluster medium. To account for the observed Faraday 
rotation, the cluster magnetic field must be present up to a distance
of more than 1 cluster core radius. 
The strength of the field is between 5 and 10 $\mu$G, 
depending on the scale of the magnetic field tangling.

\begin{acknowledgements}

The authors thank the referee, Dr. Frazer Owen, for helpful comments.

LF and GG acknowledge partial financial support from the Italian
Space Agency (ASI) and from the Italian MURST.

The NRAO is a facility of the National Science Foundation operated
under cooperative agreement by Associated Universities, Inc.

\end{acknowledgements}

%
% 20 cm. 3.5 arcsec resolution
\begin{figure*}
\label{Cont15l}
\end{figure*}
%
% 6 cm. 1.5 arcsec resolution
\begin{figure*}
\label{Cont15c}
\end{figure*}
%
% 3.6 cm. ... arcsec resolution
\begin{figure*}
\label{Cont15x}
\end{figure*}
%
% 20 cm. 3.5 arcsec resolution
\begin{figure*}
\label{Cont16l}
\end{figure*}
%
% 6 cm. 1.5 arcsec resolution
\begin{figure*}
\label{Cont16c}
\end{figure*}
%
% 3.6 cm. ... arcsec resolution
\begin{figure*}
\label{Cont16x}
\end{figure*}
%
% 6 cm. 5 arcsec resolution
\begin{figure*}
\label{Cont3c}
\end{figure*}
%
% 3.6 cm. 2 arcsec resolution
\begin{figure*}
\label{Cont3cx}
\end{figure*}
%
% radio-X overlay
\begin{figure*}
\label{Radx}
\end{figure*}
%
% RM 0053-015
\begin{figure*}
\label{rm15}
\end{figure*}
%
% Histogram RM 0053-015
\begin{figure*}
\label{rmhi15}
\end{figure*}
%
% RM 0053-016
\begin{figure*}
\label{rm16}
\end{figure*}
%
% Histogram RM 0053-016
\begin{figure*}
\label{rmhi16}
\end{figure*}
%
% RM 3c29
\begin{figure*}
\label{rm3c}
\end{figure*}
%
% Histogram RM 3c29
\begin{figure*}
\label{rmhi3c}
\end{figure*}
\end{document}